# High-pressure structures and superconductivity of bismuth hydrides

Yanbin Ma, Defang Duan, Da Li, Yunxian Liu, Fubo Tian, Hongyu Yu, Chunhong Xu, Ziji Shao,

Bingbing Liu and Tian Cui*

*State Key Laboratory of Superhard Materials, College of Physics, Jilin University, Changchun

130012, People's Republic of China*

## Abstract

We have systematically searched for the ground state structures of bismuth hydrides based on evolutionary algorithm method and particle swarm optimization algorithm method. Given only rich-hydrogen region, except for $BiH_3$, other hydrides ($BiH$, $BiH_2$, $BiH_4$, $BiH_5$, $BiH_6$) have been predicted to be stable with pressurization. With the increase of hydrogen content, hydrogen exists in bismuth hydrides with the different forms and presents the characteristics of ionicity. Under high pressure, the remarkable structural feature is the emergence of $H_2$ units in $BiH_2$, $BiH_4$ and $BiH_6$, and $BiH_5$ adopts a startling layered structure intercalated by $H_2$ and the linear $H_3$ units. Further calculations show these energetically stable hydrides are good metal and their metallic pressures are lower than that of pure solid hydrogen because of the doping impurities. The $T_c$ in the range of 20-119 K has been calculated by the Allen-Dynes modified McMillan equation, which indicates all these stable hydrides are potential high-temperature superconductors. Remarkably, it is the H-Bi-H and Bi atoms vibrations rather than the high-frequency $H_2$ or $H_3$ units that dominate the superconductivity. In addition, hydrogen content has a great influence on the superconducting transition temperature.

## Introduction

The research on superconductivity in solid state hydrogen can be traced back to 1968 [1]. However, the synthesis of metallic hydrogen as the prerequisite has been far beyond the people's expectation and the accessible experimental pressures [2, 3]. In order to circumvent the problem, it was proposed that metallization pressures could be reduced in hydrogen-rich compounds due to the impurity rendering "chemical precompression" and these hydrides will become high-temperature superconductor after metallization [4]. In the first instances, the focus was concentrated on group-IV hydrides. Explorations on high-pressure structures of silane ($SiH_4$) [5-9], disilane ($Si_2H_6$) [10, 11] germane ($GeH_4$) [12], and stannane ($SnH_4$) [13, 14] have revealed the possibility of metallic phases with high superconducting transition temperature $T_c$. These researches are not only limited in group-IV hydrides, but also other hydrides ($LiH_6$, $KH_6$, $CaH_6$, and so on) [15-17] have also been extensively explored. More recently, one quite startling report consolidates this reality in both theory and experiment. Remarkably, by using the *ab initio* calculation method, our group predicted the covalent crystal $H_3S$ [18, 19] identified with space group *Im-3m* to be an extraordinarily conventional high-temperature superconductor with a high $T_c$ values of 191 ~ 204 K at 200 GPa, which has been proved by Eremets *et al* through the high-pressure experiment [20].

More exciting, Eremets *et al* also measure a $T_c$ of another covalent hydride "phosphine" [21] (underdetermined structural information) that reaches 103 K at 207 GPa. Later, $PH_2$ [22] proposed by Andrew Shamp *et al* could be a candidate hydride according to "phosphine". The observation of superconductivity in "phosphine" activates this "stagnant water" that the superconductivity of group VA hydrides is rarely studied and it may play a guiding role on researching other hydrides in same family. For instances, the semimetal antimony (Sb) element doped into hydrogen has been performed theoretically by us to explore the high-pressure crystal structures and superconductivity. Specially, a nontrivial binding was found in $SbH_4$ [23] and it was predicted to have a high $T_c$=106 K at 150 GPa. As same family element, bismuth has a larger atomic mass which could yield low energy vibrations. Although the low-frequency vibrations may generate the low phonon frequency logarithmic average ($\omega_{log}$), they also do have the advantages in improving the electron phonon coupling (EPC). Unfortunately, Bismuth hydrides are unstable under normal pressure. As is known, pressurization is an effective way to reduce the chemical potential of the reaction and then overcome the high reaction barrier of hydrogen with some elements to promote the synthesis of hydrides.

In this letter, in order to search low-enthalpy ground-state structures in H-rich regime of Bi-H compounds, we consider enough components of bismuth-rich and hydrogen-rich compounds over a wide range of pressure (50~300 GPa). Our calculations suggest that except for $BiH_3$, all of the bismuth hydrides become stable with increasing pressure. The $H_2$ units are discovered in bismuth hydrides $BiH_2$, $BiH_4$ and $BiH_6$ by hydrogenation. A layered structure of $BiH_5$ containing $H_2$ and liner $H_3$ units is uncovered. Moreover, all the stable Bismuth hydrides are good metal containing ionic hydrogen and the superconducting critical temperature $T$c values of BiH, $BiH_2$, $BiH_4$ $BiH_5$ and $BiH_6$ are 24-30 K, 51-59 K, 81-93 K, 92-103 K and 88-100 K at 250, 150, 150, 200 and 200 GPa, respectively. The increasing hydrogen content plays a significant role in enhancement of $T_c$ for different stoichiometries hydrides.

## Computational Methods

We report on our investigations of the high-pressure crystalline structure of Bi-H system by using the evolutionary algorithm, implemented in the USPEX code [24-26] and the particle swarm optimization algorithm as implemented in CALYPSO (crystal structure analysis by particle swarm optimization) code [27, 28]. First of all, variable-composition structural predictions were executed by considering enough bismuth-rich and hydrogen-rich hydrides at 50, 100, 150, 200, 250, 300 GPa. Secondly, these most stable structures in thermodynamics were accurately searched by changing the cell size (1-4 formula units). The most promising candidate structures are further optimized with a higher accuracy by using the projector augmented waves method [29], implemented in the Vienna *ab initio* simulation package VASP code [30]. H (1s) and Bi ($6s^2 6p^3$) were treated as valence electrons. The generalized gradient approximation of Perdew-Burke-Ernzerhof [31] was chosen to describe the exchange-correlation potential. The higher accuracy includes a plane-wave basis set cutoff of 800 eV and a Brillouin zone sampling grid of spacing $2\pi \times 0.03$ Å$^{-1}$, resulting in total energy convergence better than 1 meV/atom. The phonon spectrum and electron phonon coupling (EPC) were obtained from density functional perturbation theory [32], as implemented in the QUANTUMESPRESSO code [33]. The norm-conserving pseudopotentials are used. We tested a suitable 100 Ry kinetic energy cutoff for all hydrides. For $P6_3/mmc$-BiH and $Pmmn$-BiH$_4$, a 16 ×16 × 16 Monkhorst-Pack (MP) [34] $k$-point sampling mesh is enough for the electronic Brillouin zone (BZ) integration and a $4 \times 4 \times 4$ $q$-mesh used to evaluate both the phonon band structure. A 24 ×24 × 18 MP $k$-point sampling mesh and a $4 \times 4$

× 3 $q$-mesh were employed in $P2_1/m$-BiH$_2$ and $C2/m$-BiH$_5$. A 24 ×24 × 16 MP $k$-point sampling mesh and a 4 × 4 × 2 $q$-mesh were suitable to calculate phonon dispersion curves for $P$-1-BiH$_6$.

## Results and discussion

### A. Thermodynamic stability of BiH$_n$ under pressure: from enthalpic trends

To explore which BiH$_n$ species and structures could be synthesized in principle, convex shell is constructed by relaxing the most favorable structures of BiH$_n$, illustrated in Fig 1(a). On the one hand, bismuth-rich hydrides found are unstable. On the other hand, it may be a potential rule that large hydrogen fraction may be an essential condition to produce the high $T_c$ for hydrogen-rich compounds. These bismuth-rich hydrides are not drawed on convex shell. From Fig 1(a), one can deduce which stoichiometries are thermodynamically stable at given pressures. Provided a point (corresponding to a phase) lies on convex hull and does not exhibit any imaginary phonon modes, this phase is stable. If a tie-line connect thermodynamically stable phase A with B and another phase C falls above it, it is expected that C can decompose into A and B structures. That no any unstable mode in the phonon dispersion curves of C implies it is metastable, otherwise it is unstable. The stability and structural character of bismuth hydrides depend on pressure and hydrogen environment. The essential information can be summarized from Fig. 1(a): (1) below 100 GPa, none of BiH$_n$ is stable against elemental dissociation, conforming the fact of nonexistence of Bi hydrides at low pressures; (2) at 150 GPa, both $P2_1/m$-BiH$_2$ and $Pmmn$-BiH$_4$ lie on the convex hulls, but BiH$_2$ has lower average atomic enthalpy, which indicates it could be more likely to be synthesized in experiment; (3) under further compression (when pressure is up to 200 GPa), BiH$_5$ and BiH$_6$ emerge on tie-lines, whereas BiH$_2$ is still the most stable stoicheiometry; (4) at 250 GPa, except for BiH$_3$, all stoichiometries fall on the convex shells and BiH becomes the most stable at 300 GPa. With the increasing hydrogen environment, bismuth hydrides are reasonably divided into three groups as depicted in Fig 2: (i) BiH contains hydrogen atoms; (ii) BiH$_2$, BiH$_4$ and BiH$_6$ contains H$_2$ units; (iii) BiH$_5$ contains H$_2$ units and symmetrical linear H$_3$ units.

### B. Structural characteristics with hydrogenation in BiH$_n$

The structural parameters for the BiH$_n$ structures at selected pressures are shown in Table SI (seeing Supplementary Material) and corresponding perspectives are provided in Fig. 2. The direct results from the figures reveal the commensurable impurity forms the $P6_3/mmc$-BiH where H atoms constitute a hexagonal frame with two Bi atoms occupying on 2*c* Wyckoff site. With the redundant hydrogen content, a clear structural feature in bismuth hydrides (BiH, BiH$_2$, BiH$_4$ and BiH$_6$) is appearance of H$_2$ units. For *P2$_1$/m*-BiH$_2$, it is thermodynamically stable (with respect to Bi and H$_2$) above 117 GPa and contains H$_2$ units with a H-H distance of 0.804 Å at 150 GPa. *Pmmn*-BiH$_4$ is stable from 125 to 300 GPa, where Bi atoms adopt a wrinkled layered and the H$_2$ units with a H-H distance of 0.838 Å are intercalated in adjacent Bi atoms layer. It is notable that BiH$_4$ is a unique layered structure which is different from other hydrides like KH$_6$ [17] and PbH$_4$ [35]. In BiH$_4$, only Bi atoms array as a layered and all of H$_2$ units insert into the interval of the layers, whereas the mixed H$_2$ units or H$_3$ units and K atoms constitute these layers in *C2/m*-KH$_6$. Comparing with PbH$_4$ (phase VIII), although all H$_2$ units are insert into the intervals, those H$_2$ units are coplanar in PbH$_4$, while these H$_2$ units in BiH$_4$ are intercalated intervals keeping a certain angle with XY plane (seeing Fig 2 (c)). Notably, there are two metastable BiH$_4$ identified with *P6$_3$/mmm* and *P6$_3$/mmc* symmetry also found in our work. The *P6$_3$/mmc*-BiH$_4$ structure is analogous with the previously predicted SbH$_4$ by us [23] and SnH$_4$ [14] proposed by Gao *et al.* However, it is metastable because of the higher enthalpy. When pressure reaches 200 GPa, an individual layered structure *C2/m*-BiH$_5$ containing H$_2$ and symmetrical linear H$_3$ units is uncovered, as depicted in Fig 2 (d) (e). The linear H$_3$ units are also found in other hydrides, for example CsH$_3$ [36], RbH$_5$ [37], KH$_5$ [38], BaH$_6$ [39] and Te$_2$H$_5$ [40]. Another triangular form of H$_3$ units is reported in H$_n$F [41], H$_5$Cl [41, 42] and OsH$_8$ [43]. The *C2/m*-BiH$_5$ is akin to *C2/c*-KH$_6$, however, the major difference is that H$_3$ units in KH$_6$ is unequal separation, whereas it is symmetrical (0.914 Å at 200 GPa) in BiH$_5$. Turning to the richest-H$_2$ stoichiometry BiH$_6$ (seeing Fig 2(f)), it has a low *P*-1 symmetry with a H-H distance of 0.830 Å at 200 GPa.

## C. Metallization of BiH$_n$ and the role of hydrogen

To understand the electronic properties and role of hydrogen, the electronic band structure, density of states (DOS) and Bader charges [44-46] are calculated. The overlap of electronic band near Fermi level (seeing Fig 3) implies that all stable hydrides exhibit metallic feature and the DOS near Fermi level is mainly contributed by Bi-6*p* (depicted in Fig. S2). Bismuth hydrides have been predicted to have the lower metallic pressure than that in pure solid hydrogen, supporting that impurities can reduce the metallization pressure. Specially, the calculated electron localization function (ELF) (seeing Fig. 3 and Fig. S3) suggests that the ELF value around H hydrogen is close to 0.95, which also proves the existence of H$_2$ and H$_3$ units. To explore the role of hydrogen in BiH$_n$, we calculate Bader charges listed in Table S1-S5. The obvious charges are found to transfer from Bi to H. Furthermore, we also analyzed the band structure of a hypothetical system where all Bi atoms are removed and only the H sublattices with a compensating background charge are remained. As is depicted in Fig 3(a), the band structures of $BiH_n$ and the hydrogen subsystem are qualitatively similar, which hints that Bi atoms mainly donate their electrons to hydrogen and form the ionic H. The homogeneous shape of electronic band usually indicates the ionic bonding features of Bi-H system, similar to $AlH_3$ [47], $GaH_3$ [19], $TeH_4$ [40] and $LiH_n$ [15]. However, the binding in Bi-H system is quite different from the same group hydride $SbH_4$. Our previous research on $SbH_4$ reveals, except for $H_2$ units, the weak covalent interaction was found between H and Sb. There is no doubt that the discrepancy of binding depends on both crystalline structures and elemental nature. In VA group, nitrogen (N) and phosphorus (P) are nonmetal. Arsenic and antimony are regarded as semimetal, while bismuth is the only real metal. Compared with other elements in group VA, bismuth has the heaviest atomic mass and the weakest electronegativity. In Bi-H system, H atoms may have a stronger ability in attracting electrons than Bi and then causes the ionic bonding form. Besides, energy band structures near the Fermi level for $C2/m$-$BiH_5$ is characterized as "flat band (near M point)-steep band", indicating that it may be a potential high-temperature superconductor [48].

## D. Dynamics and superconducting properties in $BiH_n$

To explore the dynamics properties of $BiH_n$, we calculated the phonon band structure. As is depicted in Fig. 4 and Fig. S4, no imaginary frequency in the phonon band structure suggests that all the researched structures are vibrationally stable. Different forms of hydrogen in Bi-H system determine the phonon band structures. For $P6_3/mmc$-BiH containing commensurable H atoms, the phonon dispersion curve is divided into two parts and the low-frequency is caused by Bi atoms vibrations whereas high-frequency is related to H atoms vibrations. For $H_2$ or $H_3$ units-contained hydrides, the phonon band structures are divided into three parts. The low-frequency and middle frequency are derived by Bi atoms vibrations and H-Bi-H vibrations, respectively, whereas the high-frequency is contributed from $H_2$ or $H_3$ units. For $BiH_5$, the phonon dispersion curve is different from others $H_2$-contained hydrides ($BiH_2$, $BiH_4$ and $BiH_6$). As is shown in Fig. 4 and Fig. S4, $BiH_2$, $BiH_4$ and $BiH_6$ have a flat moderate-frequency band, whereas for $BiH_5$, these dips appear at M and $\Gamma$, which is helpful to enhance the electron phonon coupling ($\lambda_{qv}$). Moreover, in whole Brillouin zone, all phonon modes contribute to superconductivity, which proves BiH$_5$ is an isotropic three-dimensional system.

To explore the possible superconductivity of Bi-H system at high pressures, the phonon frequency logarithmic average ($\omega_{log}$), DOS at Fermi level $N(E_f)$ and EPC parameter ($\lambda$) at stable pressures are calculated and listed in Table SII-SVI. The obtained EPC parameter $\lambda$ is 0.75 for BiH (at 250 GPa), 1.34 for BiH$_2$ (at 150GPa), 1.27 for BiH$_4$ (at 200 GPa), 1.23 for BiH$_5$ (at 200GPa) and 1.26 for BiH$_6$ (at 200 GPa), while the calculated $\omega_{log}$ from the phonon spectrum reach 699.2, 506.8, 870, 1021.9 and 934.6 K, respectively. The superconducting critical temperature $T_c$ of Bi-H system stable phases can be estimated by using the Allen-Dynes-modified McMillan equation [49] $T_C = f_1 f_2 \frac{\omega_{log}}{1.2} \exp[-\frac{1.04(1+\lambda)}{\lambda - \mu^*(1+0.62\lambda)}]$ where the coulomb pseudopotential $\mu^*$ is often taken as ~ 0.1 and 0.13. As is depicted in Table SII-SVI, with pressure increasing, the $T_c$ decreases in BiH and BiH$_4$ while increases in BiH$_2$ and BiH$_6$. For BiH$_5$, the superconducting transition temperature first increases and then decreases with increasing pressure. In addition, the $T_c$ of all hydrides increases with the increasing hydrogen-content. It is very obvious the $T_c$ have a weak dependence on pressure.

To get to the bottom of physical mechanism for superconductivity, we also analyzed phonon density of states, Eliashberg phonon spectral function $\alpha^2F(\omega)$ and the partial electron-phonon integral $\lambda$, as shown in Fig. 4 and Fig. S4. For BiH at 250 GPa (Fig. 3S (a)), the low-frequency modes (below 18 THz) vibrations contribute 47% of the total $\lambda$, while high frequency modes above 35 THz provide a contribution 53% in total EPC parameter. For the $P2_1/m$ of BiH$_2$ at 200GPa (Fig. 3S (b)), approximately 43% of total $\lambda$ is ascribed to Bi atoms vibrations (below 8.6 THz) and the 53% is from the moderate vibrations, the remaining only 4% from the H$_2$ vibrations (80-83THz). For BiH$_4$ at 150 GPa (Fig. 3S(c)), the condition is akin to $P2_1/m$-BiH$_2$. Bi atoms vibrations devote 27% to the total $\lambda$ and 66% mainly comes from H-Bi-H vibrations, whereas only 7% is from H$_2$ units. For BiH$_5$ at 200 GPa (Fig. 3), vibrations between 83 and 90 THz produced by the H$_3$ and H$_2$ units provide merely 2% to total $\lambda$, while the $\lambda$ from middle region contributes about 73% and Bi atoms contribute about 25%. For BiH$_6$, it is found that the Bi, H-Bi-H and H$_2$ units vibration modes account for 22%, 71% and 7% of $\lambda$ at 200 GPa (Figure 3S (d)). This law can be summarized that it is the H-Bi-H vibrations and low-frequency Bi atoms vibrations rather than the high-frequency H$_2$ or H$_3$ units that dominate the strength of electron-phonon coupling and superconductivity.

## Conclusion

In summary, we explore high-pressure phase and superconductivity of bismuth hydrides by using *ab initio* calculations. The bismuth hydrides (except for $BiH_3$) become thermodynamically stable with respect to decomposition into the Bi and H with increasing pressure. Except $P6_3/mmc$-BiH, a remarkable feature of the predicted stable structures is the presence of $H_2$ units or $H_3$ units. All hydrides are predicted to be potential high temperature superconductors with $T_c$ in the range of 20-119 K. The hydrogen content in bismuth hydrides plays a crucial role on impacting the $T_c$. The Superconducting transition temperature increases with the increasing hydrogen content. Our results suggest in Bi-H system the H-Bi-H and low-frequency Bi atoms vibrations contribute most significantly to the EPC. We expect our finding will open a door for studying experimental synthesis of Bi hydrides and exploring their high-pressure superconductivity.

## Acknowledgments

This work was supported by the National Basic Research Program of China (No. 2011CB808200), National Natural Science Foundation of China (Nos. 51572108, 11204100, 11574109, 11404134, 11504127), Program for Changjiang Scholars and Innovative Research Team in University (No. IRT1132), National Found for Fostering Talents of basic Science (No. J1103202). Part of calculations were performed in the High Performance Computing Center (HPCC) of Jilin University.

## References

*To whom correspondence may be addressed. Email: cuitian@jlu.edu.cn

**Captions**

Fig.1. (Color online) (a), Enthalpies of formation (with respect to Bi and $H_2$) of $BiH_n$ (n=1~6). Completely filled symbols denote the structures are on the convex shells, but for half filled represents that are not on tie-lines.

Fig.2. (Color online) (a), $P6_3/mmc$-BiH at 250 GPa. (b), (c), (d) and (f) are the YZ planes of supercells ($2 \times 2 \times 2$) for the stable structures. (b), $P2_1/m$-$BiH_2$ at 150 GPa. (c), $Pmmn$-$BiH_4$ at 150 GPa. (d), $C2/m$-$BiH_5$ at 200 GPa. (e), the XY plane maps of $C2/m$-$BiH_5$. (f), $P$-1-$BiH_6$ at 200 GPa.

Fig.3. (Color online) Calculated electronic band structure and XY plane of electron localization function (ELF) for $BiH_5$ at 200 GPa. The full black line denotes real electronic band structure and dashed red line represents the hypothetical system composed of the hydrogen sublattice with compensating background charge.

Fig.4. (Color online) (a), calculated phonon dispersion relation for $C2/m$-$BiH_5$ at 200 GPa. These cardinal circles overlapping with the phonon dispersion are proportional to the strength of EPC ($\lambda_{qv}$). (b), the phonon DOS projected on H and Bi atoms. (c) The Eliashberg phonon spectral function $\alpha^2 F(\omega)$ and the partial electron-phonon integral $\lambda(\omega)$.

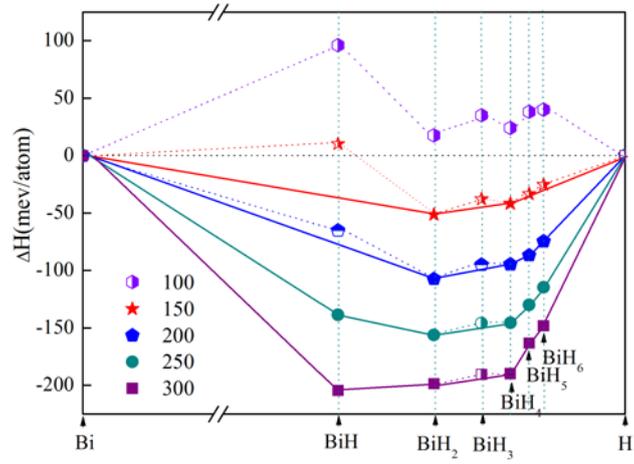

Fig. 1

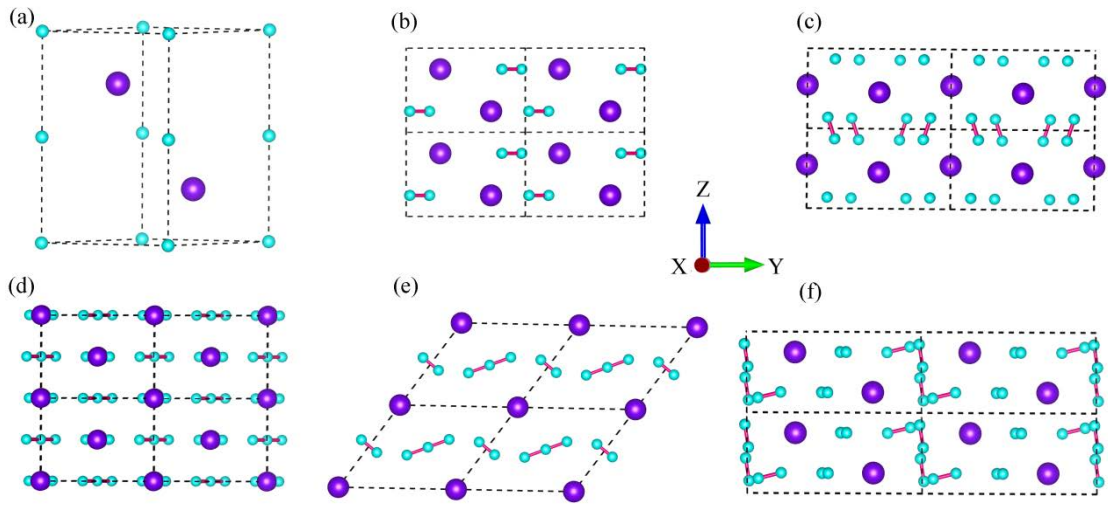

Fig. 2

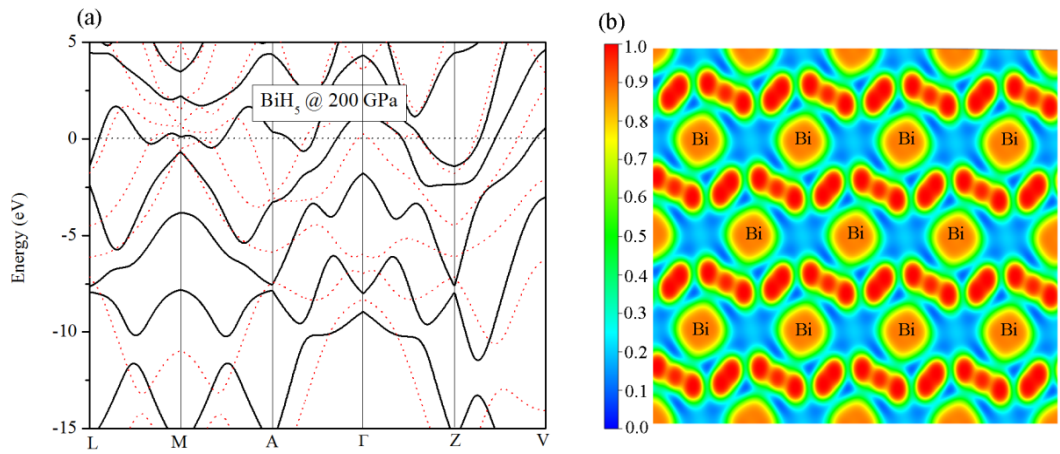

**Fig. 3**

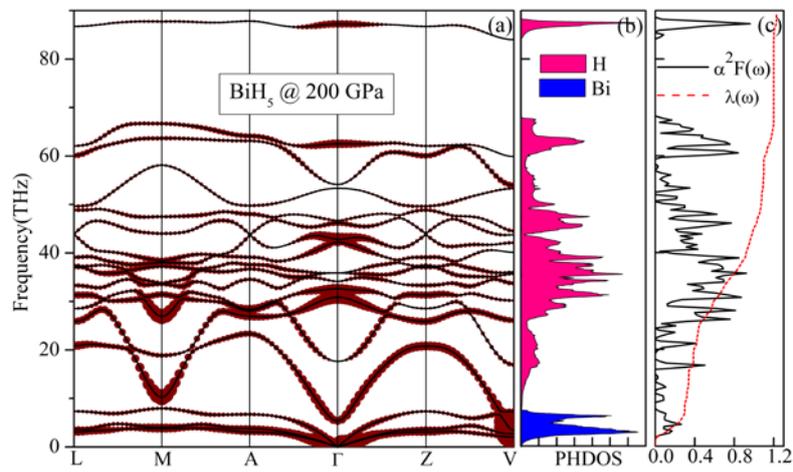

**Fig. 4**

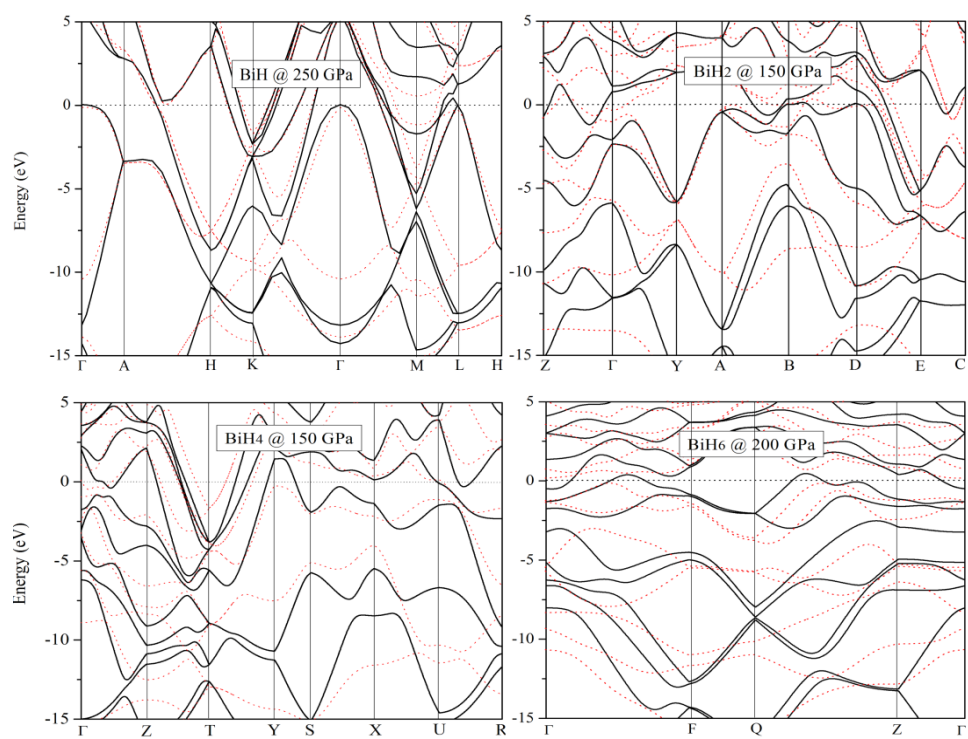

**Fig. S1** (Color online) The electronic band structure for BiH at 250, BiH$_2$ at 150, BiH$_4$ at 150 and

BiH$_6$ at 200 GPa, respectively. The full black line denotes real electronic band structure and dashed red line represents the hypothetical system composed of the hydrogen sublattice with compensating background charge.

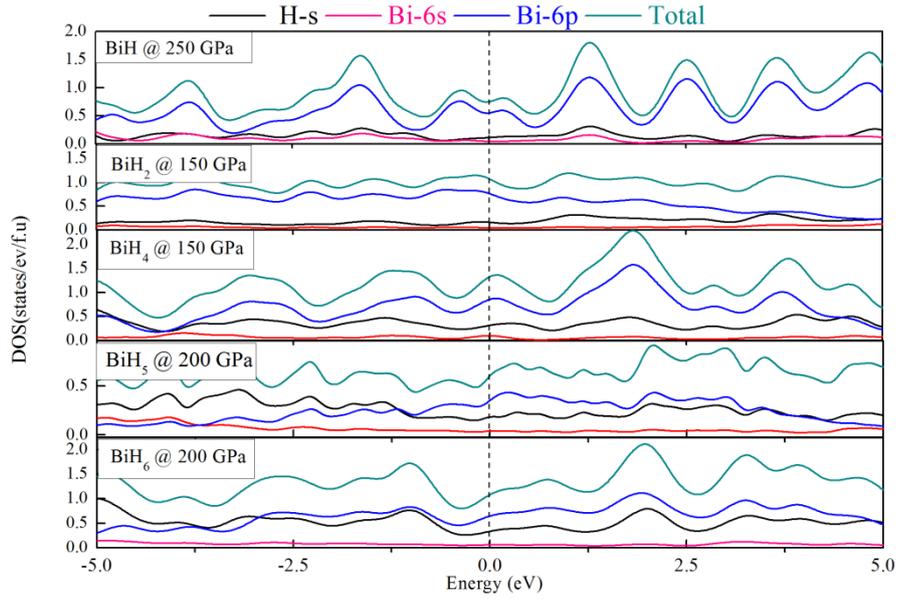

**Fig. S2** (Color online) The density of states (DOS) for BiH at 250, BiH$_2$ at 150, BiH$_4$ at 150 and BiH$_6$ at 200 GPa, respectively.

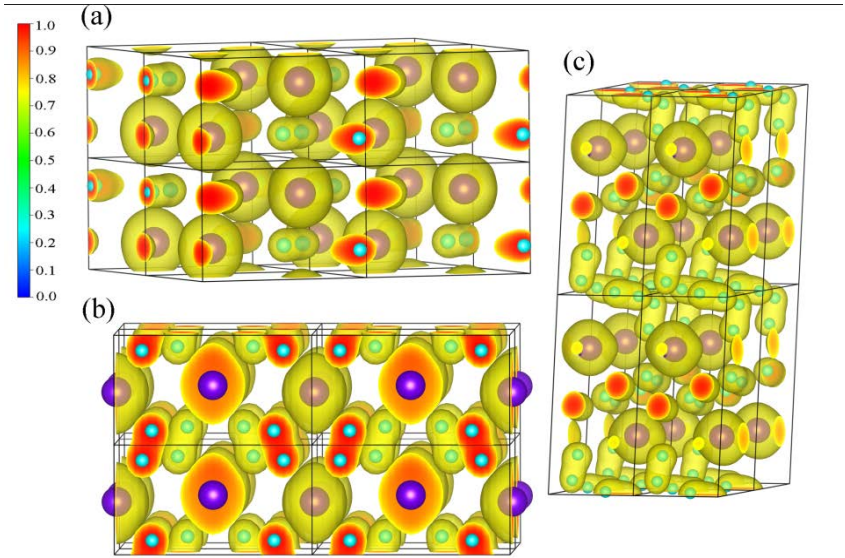

**Fig. S3** (Color online) The 3D-electron localization function (ELF) with isosurface value of 0.75. (a) $P2_1/m$-BiH$_2$ at 150 GPa, (b) $Pmmn$-BiH$_4$ at 150 GPa, (c) $P$-1-BiH$_6$ at 200 GPa.

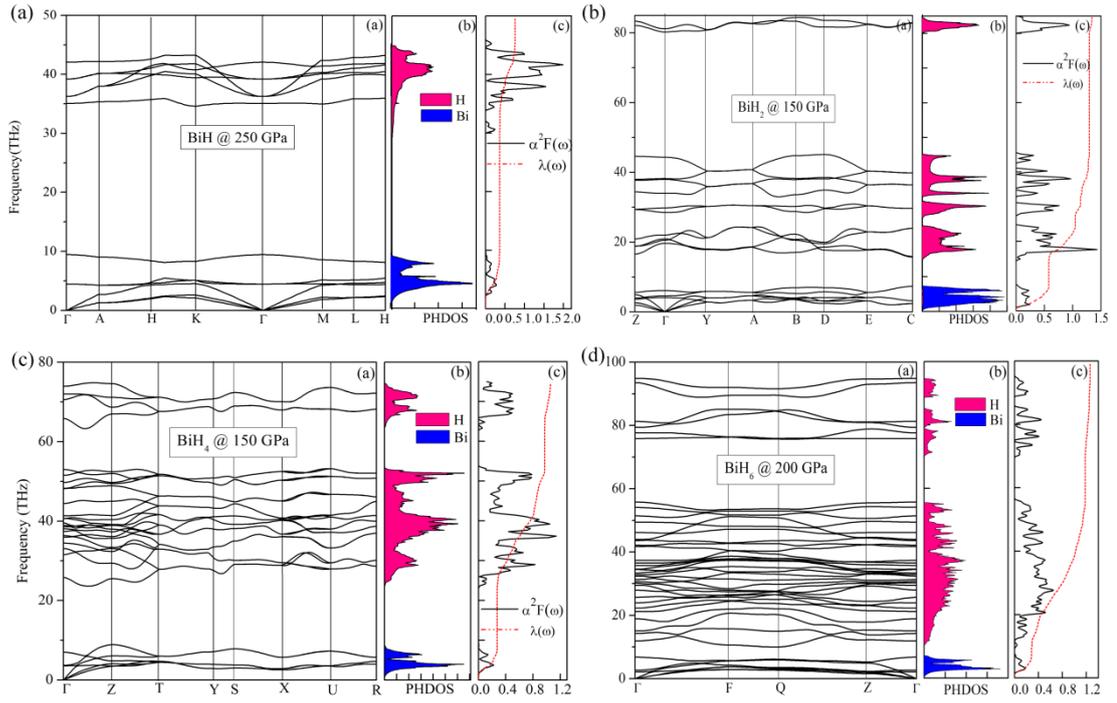

**Fig. S4.** (Color online) The phonon band structure, phonon density of states, Eliashberg phonon spectral function $\alpha^2F(\omega)$ and the partial electron-phonon integral $\lambda$ for BiH at 250, BiH$_2$ at 150, BiH$_4$ at 150 and BiH$_6$ at 200 GPa, respectively.

**Table SI**. The lattice parameters and atomic positions of $P6_3/mmc$-BiH at 250 GPa, $P2_1/m$-BiH$_2$ at 150 GPa, $Pmmn$ -BiH$_4$ at 150 GPa, $C2/m$-BiH$_5$ at 200 GPa and $P$-1-BiH$_6$ at 200GPa.

| Structure | Parameters (Å, deg) | Atom | x | y | z |
|---|---|---|---|---|---|
| BiH $P6_3/mmc$ (250 GPa) | a= 2.8355 b= 2.8355 c= 4.5234 γ= 120 | H Bi | 0.00000 0.33333 | 0.00000 0.66667 | 0.50000 0.25000 |
| BiH$_2$ $P2_1/m$ (150 GPa) | a= 4.5029 b= 3.0751 c= 3.0145 β=73.6949 | H H Bi | 0.80375 0.03102 0.28727 | 0.25000 0.75000 0.25000 | 0.15578 0.03594 0.33760 |
| BiH$_4$ $Pmmn$ (150 GPa) | a= 3.0607 b= 5.2513 c= 2.9380 | H H Bi | 0.00000 0.50000 0.00000 | 0.14251 0.30853 0.50000 | 0.13921 0.13066 0.45569 |
| BiH$_5$ $C2/m$ (200 GPa) | a= 2.7849 b= 2.7849 c= 4.1785 α=β=119.9594 γ= 61.9762 | H H H Bi | -0.37224 -0.10423 -0.50000 0.00000 | -0.37224 -0.10423 -0.50000 0.00000 | -0.59884 -0.57462 -0.50000 0.00000 |
| BiH$_6$ $P$-1 (200 GPa) | $a$= 2.8017 $b$= 2.8732 $c$= 6.2900 $α$=89.3171 $β$=79.5108 $γ$= 89.8771 | H H H H H H Bi | 0.17087 0.24640 0.87768 0.21712 0.73956 0.17570 0.63408 | 0.74618 0.43069 0.75018 0.81684 0.84885 0.75131 0.25293 | 0.19468 0.00802 0.46282 0.06420 0.01119 0.42965 0.27794 |

**Table SII.** The phonon frequency logarithmic average ($\omega_{log}$), EPC parameter ($\lambda$), the electronic DOS at Fermi level $N(E_f)$, $f_1f_2$ ($\mu$*= 0.1 and 0.13) and critical temperature $T_c$ ($\mu$*= 0.1 and 0.13) for $P6_3/mmc$-BiH at 250 and 300GPa.

| Pressure (GPa) | Lambda($\lambda$) | $\omega_{log}$(K) | $N(E_f)$ states/Spin/Ry/cell | $f_1f_2$ ($\mu$*=0.1) | $f_1f_2$ ($\mu$*=0.13) | Tc(K) ($\mu$*=0.1) | Tc(K) ($\mu$*=0.13) |
|---|---|---|---|---|---|---|---|
| 250 | 0.76 | 699.17 | 5.14 | 1.05 | 1.04 | 30 | 24 |
| 300 | 0.61 | 792.43 | 5.03 | 1.04 | 1.03 | 20 | 13 |

**Table SIII.** The phonon frequency logarithmic average ($\omega_{\log}$), EPC parameter ($\lambda$), the electronic DOS at Fermi level $N(E_f)$, $f_1f_2$ ($\mu^*= 0.1$ and 0.13) and critical temperature $T_c$ ($\mu^*= 0.1$ and 0.13) for $P2_1/m$-$BiH_2$ at 150,200, 250 and 300GPa.

| Pressure (GPa) | Lambda($\lambda$) | $\omega_{\log}$(K) | $N(E_f)$ states/Spin/Ry/cell | $f_1f_2$ ($\mu^*=0.1$) | $f_1f_2$ ($\mu^*=0.13$) | Tc(K) ($\mu^*=0.1$) | Tc(K) ($\mu^*=0.13$) |
|---|---|---|---|---|---|---|---|
| 150 | 1.34 | 506.80 | 6.74 | 1.13 | 1.11 | 59 | 51 |
| 200 | 1.07 | 706.70 | 6.53 | 1.09 | 1.08 | 60 | 52 |
| 250 | 0.99 | 841.53 | 6.45 | 1.08 | 1.07 | 63 | 52 |
| 300 | 0.96 | 917.48 | 6.41 | 1.07 | 1.06 | 65 | 55 |

**Table SIV.** The phonon frequency logarithmic average ($\omega_{\log}$), EPC parameter ($\lambda$), the electronic DOS at Fermi level $N(E_f)$, $f_1f_2$ ($\mu^*= 0.1$ and 0.13) and critical temperature $T_c$ ($\mu^*= 0.1$ and 0.13) for $Pmmn$-$BiH_4$ at 150,200, 250 and 300GPa.

| Pressure (GPa) | Lambda($\lambda$) | $\omega_{\log}$(K) | $N(E_f)$ states/Spin/Ry/cell | $f_1f_2$ ($\mu^*=0.1$) | $f_1f_2$ ($\mu^*=0.13$) | Tc(K) ($\mu^*=0.1$) | Tc(K) ($\mu^*=0.13$) |
|---|---|---|---|---|---|---|---|
| 150 | 1.27 | 864.72 | 8.01 | 1.17 | 1.10 | 93 | 81 |
| 200 | 1.10 | 1008.85 | 6.77 | 1.09 | 1.08 | 88 | 77 |
| 250 | 0.97 | 1101.09 | 5.62 | 1.08 | 1.06 | 77 | 66 |
| 300 | 0.92 | 1150.38 | 5.33 | 1.07 | 1.06 | 75 | 63 |

**Table SV.** The phonon frequency logarithmic average ($\omega_{\log}$), EPC parameter ($\lambda$), the electronic DOS at Fermi level $N(E_f)$, $f_1f_2$ ($\mu^*= 0.1$ and 0.13) and critical temperature $T_c$ ($\mu^*= 0.1$ and 0.13) for C2/m-$BiH_5$ at 200, 250 and 300GPa.

| Pressure (GPa) | Lambda($\lambda$) | $\omega_{\log}$(K) | $N(E_f)$ states/Spin/Ry/cell | $f_1f_2$ ($\mu^*=0.1$) | $f_1f_2$ ($\mu^*=0.13$) | Tc(K) ($\mu^*=0.1$) | Tc(K) ($\mu^*=0.13$) |
|---|---|---|---|---|---|---|---|
| 200 | 1.23 | 1021.90 | 4.23 | 1.11 | 1.10 | 103K | 92K |
| 250 | 1.45 | 801.68 | 4.18 | 1.15 | 1.13 | 101K | 90K |
| 300 | 1.20 | 1220.96 | 4.05 | 1.10 | 1.09 | 119k | 105K |

**Table SV.** The phonon frequency logarithmic average ($\omega_{\log}$), EPC parameter ($\lambda$), the electronic DOS at Fermi level $N(E_f)$, $f_1f_2$ ($\mu^*= 0.1$ and 0.13) and critical temperature $T_c$ ($\mu^*= 0.1$ and 0.13) for $P$-1-$BiH_6$ at 200, 250 and 300GPa.

| Pressure (GPa) | Lambda($\lambda$) | $\omega_{\log}$(K) | $N(E_f)$ states/Spin/Ry/cell | $f_1f_2$ ($\mu^*=0.1$) | $f_1f_2$ ($\mu^*=0.13$) | Tc(K) ($\mu^*=0.1$) | Tc(K) ($\mu^*=0.13$) |
|---|---|---|---|---|---|---|---|
| 200 | 1.26 | 934.62 | 7.54 | 1.12 | 1.10 | 100 | 88 |
| 250 | 1.24 | 1035.33 | 7.32 | 1.11 | 1.07 | 107 | 92 |
| 300 | 1.23 | 1102.82 | 6.76 | 1.11 | 1.09 | 113 | 100 |

**Table S1.** Calculated Bader charges of H and Bi atoms in BiH (*P6₃/mmc*) at 250 GPa.

| Pressure | Atom | Charge | σ(e) |
|----------|------|--------|------|
| 250 GPa | H1 | 1.4650 | -0.4650 |
| | H2 | 1.4650 | -0.4650 |
| | Bi1 | 4.5339 | 0.4661 |
| | Bi2 | 4.5339 | 0.4661 |

**Table S2.** Calculated Bader charges of H and Bi atoms in BiH₂ (*P2₁/m*) at 150 GPa.

| Pressure | Atom | Charge | σ(e) |
|----------|------|--------|------|
| 150 GPa | H1 | 1.1731 | -0.1731 |
| | H2 | 1.1731 | -0.1731 |
| | H3 | 1.1591 | -0.1591 |
| | H4 | 1.1591 | -0.1591 |
| | Bi1 | 4.6670 | 0.3330 |
| | Bi2 | 4.6687 | 0.3313 |

**Table S3.** Calculated Bader charges of H and Bi atoms in BiH₄ (*Pmmn*) at 150 GPa.

| Pressure | Atom | Charge | σ(e) |
|----------|------|--------|------|
| 150 GPa | H1 | 1.1490 | -0.1490 |
| | H2 | 1.1691 | -0.1691 |
| | H3 | 1.1490 | -0.1490 |
| | H4 | 1.1691 | -0.1691 |
| | H5 | 1.1789 | -0.1789 |
| | H6 | 1.1789 | -0.1789 |
| | H7 | 1.1789 | -0.1789 |
| | H8 | 1.1789 | -0.1789 |
| | Bi1 | 4.3241 | 0.6759 |
| | Bi2 | 4.3241 | 0.6759 |

**Table S4.** Calculated Bader charges of H and Bi atoms in BiH₅ (*C2/m*) at 200 GPa.

| Pressure | Atom | Charge | σ(e) |
|----------|------|--------|------|
| 200 GPa | H1 | 1.3573 | -0.3573 |
| | H2 | 1.2573 | -0.2573 |
| | H3 | 1.0973 | -0.0973 |
| | H4 | 1.1131 | -0.1131 |
| | H5 | 1.0139 | -0.0139 |
| | Bi | 4.2612 | 0.7388 |

**Table S5.** Calculated Bader charges of H and Bi atoms in $BiH_6$ (*P*-1) at 200 GPa.

| Pressure | Atom | Charge | σ(e) |
|---|---|---|---|
| 200 GPa | H1 | 1.2092 | -0.2092 |
| | H2 | 1.2092 | -0.2092 |
| | H3 | 1.0951 | -0.0951 |
| | H4 | 1.0950 | -0.0950 |
| | H5 | 1.1705 | -0.1705 |
| | H6 | 1.1705 | -0.1705 |
| | H7 | 1.0372 | -0.0372 |
| | H8 | 1.0372 | -0.0372 |
| | H9 | 1.0952 | -0.0952 |
| | H10 | 1.0952 | -0.0952 |
| | H11 | 1.1871 | -0.1871 |
| | H12 | 1.1871 | -0.1871 |
| | Bi1 | 4.2058 | 0.7942 |
| | Bi2 | 4.2058 | 0.7942 |